\pdfoutput=1
\documentclass[conference,9pt]{IEEEtran}
\IEEEoverridecommandlockouts
\usepackage{cite}
\usepackage{amsmath,amssymb,amsfonts}
\usepackage{graphicx}
\usepackage{textcomp}
\usepackage{xcolor}
\usepackage[T1]{fontenc}
\usepackage{booktabs}
\usepackage{multirow}
\usepackage{url}
\usepackage{tikz}
\usetikzlibrary{arrows.meta}

\graphicspath{{figs/}}
\newcommand{\observation}[2]{\par\vspace{2pt}\noindent\hfill{\setlength{\fboxsep}{3.5pt}\setlength{\fboxrule}{0.5pt}%
\fbox{\parbox{\dimexpr0.97\columnwidth-2\fboxsep-2\fboxrule\relax}{\textbf{Observation~#1:}\enspace#2}}}\hfill\null
\par\vspace{2pt}}

\begin{document}

\title{A Runtime System for Mitigating\\Crossbar Port Contention on SM-Partitioned GPUs}
\author{\IEEEauthorblockN{Ziwen Wang\textsuperscript{1}, Qi Luo\textsuperscript{1}, Kunlin Li\textsuperscript{1}, Changxi Liu\textsuperscript{1,*}, Dongsheng Wang\textsuperscript{2}, Yun Chen\textsuperscript{1,*}}
\IEEEauthorblockA{\textsuperscript{1}\textit{The Hong Kong University of Science and Technology (Guangzhou)}\quad
\textsuperscript{2}\textit{Tsinghua University}\\
\{zwang578, qluo615, kli082\}@connect.hkust-gz.edu.cn, \{changxiliu, yunchen\}@hkust-gz.edu.cn, wds@tsinghua.edu.cn}
\thanks{\textsuperscript{*}Corresponding authors: Changxi Liu and Yun Chen.}}
\maketitle

\begin{abstract}
To avoid scheduling delays when running concurrent workloads on a GPU, Streaming Multiprocessor (SM) partitioning mechanisms such as CUDA Green Contexts restrict each workload to separate SMs. Despite this partitioning, concurrent workloads still significantly slow down latency-critical workloads, which recent studies attribute to shared L2 cache and memory contention. However, contention also occurs at the crossbar ports connecting each Graphics Processing Cluster (GPC) of SMs to the L2 cache, as all SMs in a GPC share one port. Green Contexts spread workloads across GPCs, placing concurrent workloads in the same GPCs and forcing latency-critical memory requests to queue behind background requests. Packing workloads into separate GPCs avoids this contention, but forces latency-critical memory requests through fewer ports, limiting memory bandwidth. Consequently, neither spreading nor packing is always faster across workloads and GPUs. Based on these observations, we present \textbf{GPCPlace}, a runtime system that dynamically selects between spreading and packing by profiling workloads on the target GPU to estimate both slowdowns, and hooks kernel launches in the CUDA driver to set SM masks without application modifications. Across 14 GPUs and 112 workload cases, GPCPlace speeds up latency-critical workloads by up to $1.15\times$ on average and $2.38\times$ maximum over Green Contexts without degrading throughput.
\end{abstract}

\section{Introduction}
\label{sec:intro}

Modern applications increasingly run multiple workloads concurrently on a single GPU to improve hardware utilization. However, concurrent workloads share all Streaming Multiprocessors (SMs), so a large throughput-oriented kernel can occupy the entire GPU and delay the scheduling of latency-critical workloads. To avoid such scheduling delays, SM partitioning mechanisms, such as CUDA Green Contexts~\cite{GREENCTX}, libsmctrl~\cite{LS23}, and nvtaskset~\cite{NT25}, partition the GPU's SMs by restricting each workload to a separate subset of SMs, ensuring that latency-critical workloads always have SMs available to execute immediately. Such SM partitioning is now supported by mainstream deep learning and inference frameworks and widely adopted in LLM serving, fine-tuning, and real-time processing~\cite{MW26,DU25,BU26,HA25,BV26,MO26}.

Despite this SM partitioning, recent studies find that latency-critical workloads can still be significantly slowed down by concurrent workloads~\cite{EL25,SSV25,MO26,BV26,MW26}. These studies attribute the slowdown to contention for the shared L2 cache and GPU memory. Consequently, to mitigate this contention, existing systems adjust the number of SMs allocated to each workload~\cite{MW26,NX25,BU26,HA25,HS26,BV26}, delay background kernels to avoid concurrent execution~\cite{MO26}, or modify background kernels to limit their memory accesses~\cite{BV26}.

However, beyond the shared L2 cache and GPU memory, contention also occurs at the crossbar ports that connect each Graphics Processing Cluster (GPC) of SMs to the L2 cache. Specifically, a GPU organizes its SMs into multiple GPCs, and all SMs within a GPC share a single crossbar port to access the L2 cache. Therefore, when SMs of latency-critical and background workloads are placed in the same GPC, intensive memory requests from the background workload flood the shared crossbar port and force latency-critical requests to queue behind them, significantly slowing down the latency-critical workload. We find that CUDA Green Contexts spread each workload's SMs across multiple GPCs, which places concurrent workloads in the same GPCs and causes this contention.

To avoid this contention, latency-critical and background workloads can be packed into separate GPCs. However, packing the latency-critical workload into fewer GPCs forces its memory requests to go through fewer crossbar ports, which limits its memory bandwidth and thus slows it down. Consequently, neither spreading nor packing is always faster, and the better placement depends on which of the two slowdowns is smaller, which varies across concurrent workloads and GPUs.

Based on these observations, we present \textbf{GPCPlace}, which dynamically selects between spreading and packing a latency-critical workload to reduce its latency. Specifically, GPCPlace profiles the workloads on the target GPU to estimate both the shared-port queueing delay and the bandwidth slowdown from fewer crossbar ports, and selects the placement with the smaller slowdown. At runtime, GPCPlace hooks kernel launches in the CUDA driver and sets each kernel's SM mask to the selected GPCs, requiring no modifications to applications. Moreover, because GPCPlace keeps the number of SMs allocated to each workload unchanged, it can be directly combined with existing contention-mitigation methods.

We evaluate GPCPlace across 14 NVIDIA GPUs spanning four architectures on 112 cases of concurrent latency-critical and background workloads from real-world GPU sharing scenarios. Compared with CUDA Green Contexts, GPCPlace speeds up latency-critical workloads by up to $1.15\times$ on average and up to $2.38\times$ on individual cases without degrading total system throughput.

In summary, this paper makes the following contributions:
\begin{itemize}
    \item We uncover GPC crossbar port contention under SM partitioning and show that spreading workloads across GPCs (as in CUDA Green Contexts) causes shared-port queueing, whereas packing them into separate GPCs avoids this contention but limits memory bandwidth through fewer ports, making the better placement depend on the concurrent workloads and the GPU.
    \item We propose \textbf{GPCPlace}, a runtime system that dynamically selects between spreading and packing a latency-critical workload by profiling workloads on the target GPU to estimate both slowdowns, and transparently applies the placement in the CUDA driver without application modifications.
    \item We evaluate GPCPlace across 14 NVIDIA GPUs and 112 real-world workload cases, demonstrating up to $1.15\times$ average and $2.38\times$ maximum latency speedup over CUDA Green Contexts without throughput degradation.
\end{itemize}

\section{Background and Motivation}
\label{sec:background}

\subsection{GPU Architecture and SM Partitioning}
\label{sec:bg:structure}
\begin{figure}[t]
\centering
\includegraphics[width=\columnwidth]{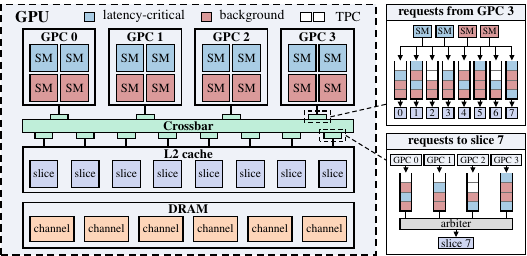}
\caption{Architecture of a GPU shared by two concurrent workloads, with one input port and one output port of the crossbar enlarged.}
\label{fig:structure}
\end{figure}

Fig.~\ref{fig:structure} shows the architecture of an NVIDIA GPU. Its compute resources are organized into multiple Graphics Processing Clusters (GPCs), each containing several Texture Processing Clusters (TPCs) of two Streaming Multiprocessors (SMs). All SMs share an on-chip L2 cache consisting of multiple slices and off-chip DRAM consisting of multiple channels. An on-chip crossbar connects the GPCs to the L2 slices, with one crossbar port for each GPC and one for each L2 slice, through which SMs send memory requests across all L2 slices. Because each GPC port sends at most one request per cycle and a busy L2 slice cannot accept every request immediately, each GPC port maintains a separate queue for each L2 slice to hold pending requests from the SMs in that GPC. When multiple GPC ports have pending requests for the same L2 slice, an arbiter selects requests equally across their queues and forwards at most one request per cycle to that slice.

SM partitioning mechanisms, such as the most widely used CUDA Green Contexts~\cite{GREENCTX}, partition a GPU's SMs across concurrent workloads by restricting each workload's kernels to a disjoint set of SMs. Specifically, these mechanisms allocate SMs at the granularity of TPCs. For example, when a latency-critical workload runs concurrently with a background workload, each workload is assigned to a separate set of TPCs so that they do not contend for compute resources (Fig.~\ref{fig:structure}). However, even with SM partitioning, concurrent workloads still contend for the shared L2 slices and DRAM~\cite{EL25,SSV25,MO26,BV26,MW26,HS26}. Moreover, because all TPCs in a GPC share one crossbar port, workloads placed in the same GPC also contend for that GPC's crossbar port, so whether this contention occurs depends on how their TPCs are assigned across GPCs.

\subsection{Green Contexts' Spread Placement Causes Requests to Queue at Shared Ports}
\label{sec:bg:queue}
\begin{figure}[t]
\centering
\includegraphics[width=\columnwidth]{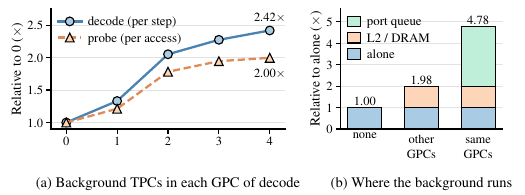}
\caption{Time of decoding and of a probe's memory accesses as background TPCs join decoding's GPCs (a), and decoding
time relative to running alone (b).}
\label{fig:shared_port}
\end{figure}

When creating a Green Context, an application specifies only the number of SMs it needs, leaving the CUDA driver to decide which TPCs across the GPCs to assign. To examine how the CUDA driver assigns TPCs across GPCs, we create Green Contexts with different SM counts across 14 NVIDIA GPUs spanning four architectures (Section~\ref{sec:eval}) and record the assigned TPCs. Across all 14 GPUs, instead of filling one GPC before moving to the next, the driver spreads each workload across multiple GPCs. Because this spread placement leaves remaining TPCs in each occupied GPC, when a latency-critical workload runs concurrently with a background workload under Green Contexts, the driver assigns those remaining TPCs to the background workload and places both workloads in the same GPCs.

However, because all TPCs in a GPC send their memory requests through the same crossbar port and its queues (Section~\ref{sec:bg:structure}), placing latency-critical and background workloads in the same GPCs causes latency-critical memory requests to queue behind background requests at each shared port, even though they do not share any SMs. To evaluate the impact of sharing GPCs, we run a latency-critical LLM decoding workload (spread across GPCs with 1~TPC per GPC) concurrently with a memory-intensive background vector addition workload on RTX~4090, and vary the number of background TPCs sharing each GPC with decoding from 0 to 4 while keeping the total number of background TPCs on the GPU constant. To quantify the queueing delay at the shared crossbar port, we place a single-thread pointer-chase probe kernel in the same GPC as the decoding workload and measure the latency of each memory access sent through that port. As shown in Fig.~\ref{fig:shared_port}a, as the number of background TPCs per shared GPC increases from 0 to 4, the probe's memory access latency increases from 193\,ns to 386\,ns ($2.00\times$) and each decoding step slows down from 65.9\,ms to 159.3\,ms ($2.42\times$). Moreover, as shown in Fig.~\ref{fig:shared_port}b, compared with running alone (33.3\,ms per step), decoding slows down by $1.98\times$ when the background workload runs in other GPCs due to shared L2 cache and DRAM contention, and by $4.78\times$ when sharing GPCs with 4 background TPCs per GPC, showing that shared-port queueing further slows down the workload beyond L2 cache and DRAM contention.

\observation{1}{Green Contexts spread workloads across GPCs, placing latency-critical and background workloads in the same GPCs and causing latency-critical memory requests to queue behind background requests at shared crossbar ports.}

\subsection{Packing Workloads into Dedicated GPCs Limits Bandwidth Through Fewer Ports}
\label{sec:bg:pack}
\begin{figure}[t]
\centering
\includegraphics[width=\columnwidth]{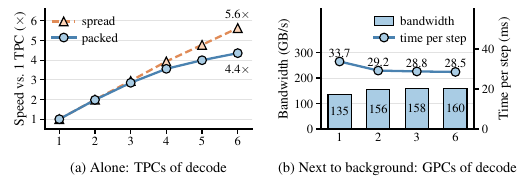}
\caption{Decoding alone, packed into one GPC or spread across GPCs (a), and next to a background workload with its
six TPCs in 1 to 6 GPCs (b).}
\label{fig:packing}
\end{figure}

To avoid queueing at shared crossbar ports (Section~\ref{sec:bg:queue}), a latency-critical workload can be isolated from background workloads by packing its TPCs into dedicated GPCs. However, doing so forces its memory requests through fewer crossbar ports, which limits its memory bandwidth and slows down the workload. To evaluate this limitation, we run the same LLM decoding workload alone on RTX~4090, and compare packing 1 to 6~TPCs into a single GPC with spreading them across 1 to 6~GPCs at 1~TPC per GPC. As shown in Fig.~\ref{fig:packing}a, packing and spreading achieve the same speed at 1 to 2~TPCs, but as the number of TPCs increases from 3 to 6, the speedup of the packed workload falls behind the near-linear scaling of the spread workload, reaching only $4.4\times$ over 1~TPC at 6~TPCs compared with $5.6\times$ when spread, confirming that restricting the workload to fewer crossbar ports slows it down by limiting its memory bandwidth~\cite{NOC24}.

Moreover, when a latency-critical workload runs concurrently with a background workload and multiple GPC ports have pending requests for the same L2 slice, the slice's arbiter selects requests equally across these ports (Section~\ref{sec:bg:structure}). Consequently, packing the latency-critical workload into fewer dedicated GPCs gives it fewer ports in this arbitration, reducing the memory bandwidth it receives from the L2 slices and slowing it down. To evaluate this limitation, we run the 6-TPC LLM decoding workload concurrently with the background vector addition workload placed in other GPCs on RTX~4090, and distribute the decoding workload's 6~TPCs across 1, 2, 3, or 6 dedicated GPCs. As shown in Fig.~\ref{fig:packing}b, when the decoding workload is packed into 1 dedicated GPC, it receives only 135\,GB/s of memory bandwidth and takes 33.7\,ms per step, whereas distributing its 6~TPCs across 2 to 6 dedicated GPCs increases its bandwidth to 156--160\,GB/s and reduces each step to 29.2--28.5\,ms.

\observation{2}{Packing a latency-critical workload into dedicated GPCs avoids shared-port queueing, but restricts the workload to fewer crossbar ports, limiting its memory bandwidth through these ports and slowing down the workload.}

\subsection{Whether Spreading or Packing Is Faster Depends on the Workloads and the GPU}
\label{sec:bg:depends}
\begin{figure}[t]
\centering
\includegraphics[width=\columnwidth]{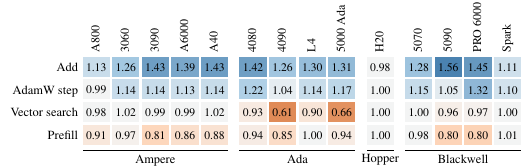}
\caption{Speedup of packing LLM decoding into dedicated GPCs over spreading it across GPCs, next to four background
workloads on 14 GPUs (above 1, in blue: packing is faster).}
\label{fig:packed_vs_spread}
\end{figure}

Spreading a latency-critical workload across GPCs causes its requests to queue at shared ports (Section~\ref{sec:bg:queue}), whereas packing it into dedicated GPCs limits its memory bandwidth through fewer ports (Section~\ref{sec:bg:pack}). Consequently, neither spreading nor packing is always faster for the latency-critical workload, and which placement achieves lower latency depends on both the concurrent workloads and the GPU. To evaluate how this trade-off varies across workloads and GPUs, we run the LLM decoding workload concurrently with four background workloads (vector addition, an AdamW optimizer step, vector search, and LLM prefill) across 14 GPUs, and measure the speedup of packing decoding into dedicated GPCs over spreading it across GPCs (Fig.~\ref{fig:packed_vs_spread}). As shown across the four rows of Fig.~\ref{fig:packed_vs_spread}, which placement is faster depends strongly on the concurrent workloads. With memory-intensive \emph{Add} or \emph{AdamW step}, which generate heavy DRAM traffic and cause severe queueing at shared crossbar ports (Section~\ref{sec:bg:queue}), packing avoids this queueing and speeds up decoding over spreading on nearly all GPUs ($1.11\times$--$1.56\times$ with \emph{Add} on 13 GPUs and $1.04\times$--$1.32\times$ with \emph{AdamW step} on 12 GPUs). Conversely, \emph{Prefill} is compute-bound with minimal memory traffic, so spreading causes negligible shared-port queueing while packing limits decoding's memory bandwidth through fewer ports (Fig.~\ref{fig:packing}a), slowing it down on most GPUs (down to $0.80\times$--$0.81\times$ on RTX~3090, RTX~5090, and RTX~PRO~6000). Similarly, \emph{Vector search} keeps the L2 slices busy without saturating individual GPC ports, so packing gives decoding fewer ports in L2 slice arbitration (Fig.~\ref{fig:packing}b) and slows it down to $0.61\times$--$0.66\times$ on RTX~4090 and RTX~5000~Ada.

Moreover, as shown across the 14 columns of Fig.~\ref{fig:packed_vs_spread}, even for the same concurrent workloads, the effect of packing versus spreading varies significantly across GPUs. For example, with \emph{Add} and \emph{AdamW step}, packing provides little speedup on H20 and A800 ($0.98\times$--$1.13\times$) due to their high HBM bandwidth, but speeds up decoding by up to $1.45\times$--$1.56\times$ on RTX~PRO~6000 and RTX~5090. Similarly, with \emph{Vector search} and \emph{Prefill}, packing slows down decoding to $0.61\times$--$0.80\times$ on RTX~4090, RTX~5000~Ada, and RTX~5090, while having little effect on H20 and DGX~Spark ($0.98\times$--$1.01\times$). Overall, the difference between packing and spreading across GPUs roughly increases with the ratio of compute request rate ($\text{SMs} \times \text{clock frequency}$) to memory bandwidth.

\observation{3}{Neither spreading nor packing is always faster for a latency-critical workload: which placement achieves lower latency depends on both the concurrent workloads and the GPU.}

\begin{figure*}[t]
\centering
\includegraphics[width=\textwidth]{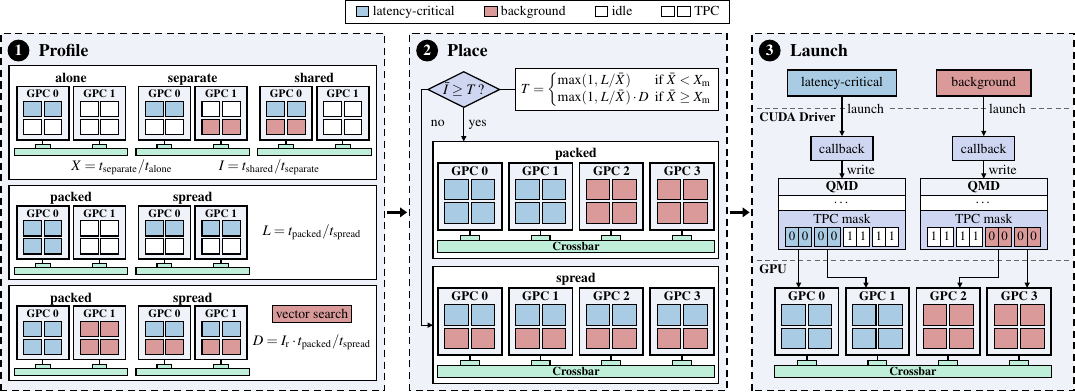}
\caption{Overview of GPCPlace: it profiles the workloads (\textcircled{1}), selects between spreading and packing (\textcircled{2}), and sets the TPC mask on each kernel launch (\textcircled{3}).}
\label{fig:overview}
\end{figure*}

\section{GPCPlace}
\label{sec:shape}

\subsection{Overview}
\label{sec:shape:overview}
Based on the observations in Section~\ref{sec:background}, we design GPCPlace, a runtime system that keeps each workload's SM count unchanged and dynamically selects between spreading and packing a latency-critical workload to reduce its latency. As shown in Fig.~\ref{fig:overview}, GPCPlace operates in three steps. First, before deployment, GPCPlace profiles the workloads on the target GPU (\textcircled{1}) to estimate the shared-port queueing delay caused by concurrent background workloads (Observation~1) and the bandwidth slowdown from packing the latency-critical workload into fewer GPCs (Observation~2). Second, GPCPlace packs the latency-critical workload into dedicated GPCs if the estimated queueing delay exceeds the packing slowdown, or spreads it across GPCs otherwise (Observation~3), and assigns the remaining TPCs to the background workloads (\textcircled{2}). Third, at runtime, GPCPlace hooks kernel launches in the CUDA driver and sets each kernel's SM mask to the selected GPCs (\textcircled{3}), requiring no modifications to applications.

\subsection{Profiling and Placement Policy}
\label{sec:shape:lat}
\label{sec:shape:decide}
\label{sec:shape:profile}
When a latency-critical workload is spread across GPCs, its memory requests queue behind background requests at shared crossbar ports (Observation~1). To estimate this queueing delay, GPCPlace profiles the latency-critical workload on 1~TPC under three settings for each background workload (Fig.~\ref{fig:overview} \textcircled{1}, top): (i)~\emph{alone}, where the latency-critical workload runs alone on 1~TPC with latency $t_{\mathrm{alone}}$; (ii)~\emph{separate}, where the background workload runs on $G - 1$~TPCs in another GPC (where $G$ is the number of TPCs per GPC) while the latency-critical workload runs with latency $t_{\mathrm{separate}}$; and (iii)~\emph{shared}, where the background workload runs on the other $G - 1$~TPCs in the same GPC as the latency-critical workload, which runs with latency $t_{\mathrm{shared}}$. Because the \emph{shared} and \emph{separate} runs have the same $G - 1$ background TPCs contending for the L2 cache and DRAM and differ only in whether they share a crossbar port, GPCPlace estimates the shared-port queueing delay avoided by packing as
\[
I = \frac{t_{\mathrm{shared}}}{t_{\mathrm{separate}}}.
\]
Meanwhile, GPCPlace also measures how much the background workload slows down the latency-critical workload through the shared L2 slices and DRAM:
\[
X = \frac{t_{\mathrm{separate}}}{t_{\mathrm{alone}}}.
\]
When multiple background workloads run concurrently, GPCPlace averages $I$ and $X$ across them into $\bar{I}$ and $\bar{X}$, weighting each background workload by its number of TPCs because a spread latency-critical workload shares more GPCs with a background workload that occupies more TPCs.

Packing a latency-critical workload into dedicated GPCs restricts it to fewer crossbar ports (Observation~2), which slows down the workload due to both limited port bandwidth (Fig.~\ref{fig:packing}a) and reduced arbitration share at busy L2 slices (Fig.~\ref{fig:packing}b). First, to measure the slowdown from limited port bandwidth, GPCPlace runs the latency-critical workload with its allocated TPCs alone under packed and spread placements (Fig.~\ref{fig:overview} \textcircled{1}, middle), measures its latencies $t_{\mathrm{packed}}$ and $t_{\mathrm{spread}}$, and computes the standalone packing slowdown $L$:
\[
L = \frac{t_{\mathrm{packed}}}{t_{\mathrm{spread}}}.
\]
When running concurrently with background workloads, a spread workload is slowed down by $\bar{X}$ at the shared L2 slices and DRAM. Meanwhile, a packed workload is bounded by the tighter bottleneck between its fewer crossbar ports ($L$) and the shared L2 slices and DRAM ($\bar{X}$), so it is slowed down by $\max(L, \bar{X})$. Dividing the two gives the port-bandwidth slowdown of packing over spreading under concurrent execution, $\max(L, \bar{X}) / \bar{X} = \max(1, L / \bar{X})$.

Second, to measure the slowdown $D$ from reduced arbitration share at busy L2 slices, GPCPlace runs the latency-critical workload with its allocated TPCs under packed and spread placements while a reference background workload occupies the remaining TPCs, and measures its latencies $t_{\mathrm{packed}}$ and $t_{\mathrm{spread}}$ under this reference workload (Fig.~\ref{fig:overview} \textcircled{1}, bottom). In general, the ratio $t_{\mathrm{packed}} / t_{\mathrm{spread}}$ under a concurrent background workload also reflects the two effects above: shared-port queueing, which slows down the spread run, and limited port bandwidth, which slows down the packed run. We use vector search as the reference background workload because under it almost only $D$ remains: it keeps the L2 slices busy, yet its $X$ value ($X_{\mathrm{r}}$) satisfies $X_{\mathrm{r}} > L$ so limited port bandwidth causes no slowdown, and its $I$ value ($I_{\mathrm{r}}$) is near $1$ ($I_{\mathrm{r}} \leq 1.03$) so shared-port queueing is minimal. GPCPlace therefore multiplies $t_{\mathrm{packed}} / t_{\mathrm{spread}}$ by $I_{\mathrm{r}}$ to remove the small shared-port queueing in the spread run and obtain $D$:
\[
D = I_{\mathrm{r}} \cdot \frac{t_{\mathrm{packed}}}{t_{\mathrm{spread}}}.
\]
Because GPU workloads execute the same kernels across iterations, GPCPlace profiles only a few iterations of each workload.

Finally, GPCPlace packs the latency-critical workload into dedicated GPCs if the shared-port queueing delay exceeds the total packing slowdown, $\bar{I} \geq T$, and spreads it across GPCs otherwise (Fig.~\ref{fig:overview} \textcircled{2}), where
\[
T = \begin{cases}
\max\bigl(1,\, L / \bar{X}\bigr) & \text{if } \bar{X} < X_{\mathrm{m}},\\
\max\bigl(1,\, L / \bar{X}\bigr) \cdot D & \text{if } \bar{X} \geq X_{\mathrm{m}}.
\end{cases}
\]
The arbitration slowdown $D$ occurs only when the background workloads keep the L2 slices busy. To identify whether the L2 slices are busy, GPCPlace uses two reference background workloads: vector search, which keeps the L2 slices busy, and LLM prefill, which does not. GPCPlace includes $D$ in $T$ only when $\bar{X} \geq X_{\mathrm{m}}$, where $X_{\mathrm{m}}$ is the average $X$ value of these two reference workloads. After placing the latency-critical workload, GPCPlace assigns the remaining TPCs to the background workloads.

\subsection{Runtime Placement}
\label{sec:shape:launch}
\begin{figure*}[t]
\centering
\includegraphics[width=\textwidth]{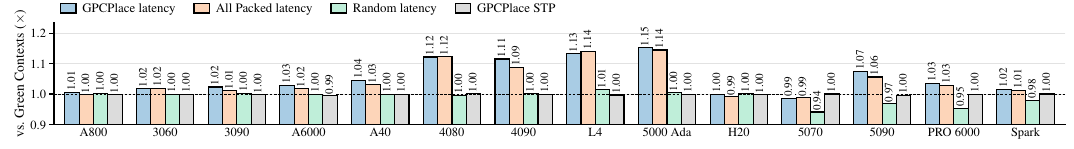}
\caption{Latency speedup of GPCPlace, All Packed, and Random over Green Contexts on each GPU, and GPCPlace's STP.}
\label{fig:eval}
\end{figure*}
Although CUDA Green Contexts partition SMs across workloads, they allow specifying only the number of SMs and provide no interface to select which GPCs or TPCs are assigned (Section~\ref{sec:bg:queue}). To implement the placement determined in Section~\ref{sec:shape:decide}, GPCPlace leverages the Queue Metadata Descriptor (QMD)~\cite{LO25,OGD} in the CUDA driver. When launching a kernel, the CUDA driver populates a QMD to describe the kernel's execution configuration for the GPU hardware. This QMD contains a TPC mask with one bit per TPC, and the hardware dispatches the kernel's thread blocks only to the TPCs enabled by this mask~\cite{LS23}. Following prior work~\cite{LS23}, at runtime, GPCPlace registers a launch callback in the CUDA driver that hooks each kernel launch before its QMD is submitted and sets the QMD's TPC mask to the TPCs assigned to the launching workload (Fig.~\ref{fig:overview} \textcircled{3}), requiring no modifications to applications.

However, NVIDIA does not document the mapping from the bits in the TPC mask to the physical TPCs on the GPU, leaving it unknown which GPC each bit controls, and prior work~\cite{LS23} only partitions TPCs by count without identifying their GPCs. Therefore, on each target GPU, we test each bit in the TPC mask by disabling only that bit, launching a small kernel that reads the SM and GPC IDs from hardware special registers, and identifying which TPC and GPC the kernel no longer executes on. Repeating this across all bits yields the mapping from TPC mask bits to GPCs on each target GPU, which GPCPlace uses to implement the selected placement at runtime.

\section{Evaluation}
\label{sec:eval}

\subsection{Experimental Setup}
\label{sec:eval:setup}

\begin{table}[t]
\centering
\caption{Specifications of the 14 evaluated NVIDIA GPUs, ordered by SMs\,$\times$\,clock / memory bandwidth within each
architecture.}
\label{tab:gpus}
\footnotesize
\setlength{\tabcolsep}{1.85pt}
\begin{tabular}{@{}llccccl@{}}
\toprule
\multirow{2}{*}{GPU} & \multirow{2}{*}{Arch.} & \multirow{2}{*}{GPCs} & \multirow{2}{*}{TPCs/GPC} & \multirow{2}{*}{SMs}
  & Clock & Memory \\
    &       &      &          &     & (GHz) & (GB/s) \\
\midrule
A800          & Ampere    & 7  & 7--8 & 108 & 1.41 & 2039\,/\,HBM2e   \\
RTX~3060      & Ampere    & 3  & 4--5 & 28  & 1.78 & 360\,/\,GDDR6   \\
RTX~3090      & Ampere    & 7  & 5--6 & 82  & 1.70 & 936\,/\,GDDR6X  \\
RTX~A6000     & Ampere    & 7  & 6    & 84  & 1.80 & 768\,/\,GDDR6   \\
A40           & Ampere    & 7  & 6    & 84  & 1.74 & 696\,/\,GDDR6   \\
\midrule
RTX~4080      & Ada       & 7  & 4--6 & 76  & 2.51 & 717\,/\,GDDR6X  \\
RTX~4090      & Ada       & 11 & 5--6 & 128 & 2.52 & 1008\,/\,GDDR6X  \\
L4            & Ada       & 5  & 5--6 & 58  & 2.04 & 300\,/\,GDDR6   \\
RTX~5000~Ada  & Ada       & 10 & 5    & 100 & 2.55 & 576\,/\,GDDR6   \\
\midrule
H20           & Hopper    & 7  & 5--6 & 78  & 1.98 & 4000\,/\,HBM3    \\
\midrule
RTX~5070      & Blackwell & 5  & 4--5 & 48  & 2.51 & 672\,/\,GDDR7   \\
RTX~5090      & Blackwell & 11 & 7--8 & 170 & 2.41 & 1792\,/\,GDDR7   \\
RTX~PRO~6000  & Blackwell & 12 & 7--8 & 188 & 2.43 & 1597\,/\,GDDR7   \\
DGX~Spark     & Blackwell & 4  & 6    & 48  & 2.42 & 273\,/\,LPDDR5x \\
\bottomrule
\end{tabular}
\end{table}
\textbf{Platforms.} We evaluate GPCPlace across 14 NVIDIA GPUs spanning the Ampere, Ada, Hopper, and Blackwell architectures, covering consumer, workstation, and datacenter GPUs (Table~\ref{tab:gpus}).

\textbf{Workloads.} To cover common GPU sharing scenarios, we construct and evaluate 112 cases of concurrent latency-critical and background workloads on each GPU. We use 4 latency-critical workloads: LLM decoding (8 Llama-3-8B layers with Triton or cuBLAS GEMMs, and attention-only decoding with FlashAttention, all at batch size 16 and context length 2048) and an embedding lookup followed by a feed-forward network. We pair them with 12 background workloads from prior studies: LLM prefill~\cite{MW26,CF26,SGL25}, an AdamW optimizer step~\cite{OR24,TA25,HA25}, a BERT-large encoder layer~\cite{SS25,HS26}, large-batch decoding~\cite{MS24}, vector search~\cite{VL26}, three cuBLAS/Triton GEMMs, vector addition, row-wise sum, matrix transpose, and GEMV. On each GPU, we allocate TPC counts equal to 1 and 3 GPCs to the latency-critical workload (with the background workload occupying the remaining TPCs), plus 4 cases with two concurrent background workloads at the 1-GPC count.

\textbf{Baselines and metrics.} We compare GPCPlace with three baselines: (i)~\emph{Green Contexts}~\cite{GREENCTX}, which spread each workload across multiple GPCs; (ii)~\emph{All Packed}~\cite{NT25}, which packs each workload into dedicated GPCs; and (iii)~\emph{Random}, which assigns TPCs to each workload randomly. To evaluate how effectively GPCPlace reduces the latency of latency-critical workloads without degrading total system throughput, we report two metrics: (1)~the execution latency of the latency-critical workload, and (2)~the total system throughput $\mathrm{STP} = \sum_i r_i^{\mathrm{s}}/r_i^{\mathrm{a}}$~\cite{SMK16} across all concurrent workloads, where $r_i^{\mathrm{s}}$ and $r_i^{\mathrm{a}}$ are the throughput of workload $i$ when running concurrently and when running alone. All results are the median of five runs.

\subsection{Overall Performance}
\label{sec:eval:lat}

Fig.~\ref{fig:eval} shows the latency speedup of GPCPlace over Green Contexts across all 112 cases on each GPU. Across the 14 GPUs, GPCPlace speeds up latency-critical workloads by up to $1.15\times$ on average and up to $2.38\times$ on individual cases over Green Contexts. Consistent with Section~\ref{sec:bg:depends}, GPUs with a higher ratio of compute request rate ($\text{SMs} \times \text{clock frequency}$) to memory bandwidth experience stronger shared-port queueing under Green Contexts and achieve larger average speedups under GPCPlace: $1.11\times$--$1.15\times$ across the four Ada GPUs, up to $1.07\times$ on the Blackwell GPUs, and $1.02\times$--$1.04\times$ across the four GDDR Ampere GPUs, compared with $1.00\times$--$1.01\times$ on the high-bandwidth HBM GPUs (A800 and H20). Meanwhile, across all 14 GPUs, GPCPlace maintains a median STP of $1.00\times$ over Green Contexts, confirming that reducing latency does not degrade total system throughput.

Fig.~\ref{fig:eval} also compares GPCPlace with \emph{All Packed} and \emph{Random}. Because \emph{All Packed} always packs every workload into dedicated GPCs, it suffers from limited port bandwidth and reduced L2 slice arbitration share when running concurrently with compute-bound or L2-saturating background workloads, slowing down the latency-critical workload to as low as $0.58\times$ of Green Contexts on RTX~4090 and $0.66\times$ on RTX~5000~Ada. By spreading the latency-critical workload when the packing slowdown exceeds the shared-port queueing delay ($\bar{I} < T$), GPCPlace avoids these slowdowns and speeds up the workload by up to $1.68\times$ over \emph{All Packed} on RTX~4090, $1.52\times$ on RTX~5000~Ada, $1.48\times$ on A40, and $1.34\times$ on RTX~PRO~6000. Similarly, because \emph{Random} assigns TPCs across GPCs randomly, it fails to isolate the latency-critical workload from shared-port queueing, allowing GPCPlace to speed up the workload over \emph{Random} by up to $1.15\times$ on average ($1.11\times$--$1.15\times$ on the Ada GPUs and RTX~5090) and up to $2.32\times$ on individual cases.

\begin{figure}[t]
\centering
\includegraphics[width=\columnwidth]{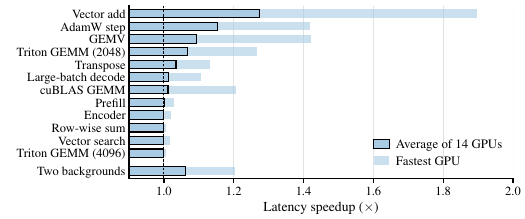}
\caption{Latency speedup of GPCPlace over Green Contexts for each background workload.}
\label{fig:eval_bg}
\end{figure}

Fig.~\ref{fig:eval_bg} shows the latency speedup of GPCPlace over Green Contexts for each background workload. When running concurrently with memory-intensive background workloads such as vector addition, the AdamW step, and GEMV, the latency-critical workload experiences severe shared-port queueing under Green Contexts, so GPCPlace avoids this queueing by packing the workload into dedicated GPCs, achieving average speedups of $1.09\times$--$1.27\times$ across the 14 GPUs and up to $1.42\times$--$1.90\times$ on the fastest GPU. Conversely, when running concurrently with compute-bound workloads (such as LLM prefill, the encoder layer, and large GEMMs) or L2-saturating vector search, where packing slows down the latency-critical workload due to limited port bandwidth and reduced L2 slice arbitration share, GPCPlace spreads the workload across GPCs and matches Green Contexts ($1.00\times$).

\subsection{Ablation Study}
\label{sec:eval:thr}

To evaluate the contribution of each component in GPCPlace's placement policy (Section~\ref{sec:shape:decide}), we compare full GPCPlace with three ablated variants of the packing slowdown $T$ across the cases where each component changes the placement: (1)~ignoring the slowdown from limited port bandwidth ($\max(1, L/\bar{X})$); (2)~ignoring the slowdown from reduced arbitration share at busy L2 slices ($D$); and (3)~always applying $D$ without checking whether the L2 slices are busy ($\bar{X} \geq X_{\mathrm{m}}$).

First, without $\max(1, L/\bar{X})$, GPCPlace underestimates the packing slowdown when running concurrently with compute-bound background workloads (such as LLM prefill and the encoder layer) and packs the latency-critical workload into dedicated GPCs even when fewer ports restrict its bandwidth (Fig.~\ref{fig:packing}a), slowing it down to an average of $0.83\times$ of Green Contexts on RTX~3090 and $0.90\times$ on RTX~4090 and RTX~5090 where a single GPC limits port bandwidth, whereas full GPCPlace spreads the workload ($0.99\times$--$1.01\times$). Second, omitting $D$ causes GPCPlace to pack the latency-critical workload when running concurrently with L2-saturating background workloads such as vector search, where fewer ports reduce the bandwidth it receives from busy L2 slices (Fig.~\ref{fig:packing}b). On GPUs where vector search heavily saturates the L2 slices (such as A40, RTX~4090, and RTX~5000~Ada), this variant slows down the workload to $0.80\times$, $0.81\times$, and $0.89\times$ of Green Contexts on average, compared with $0.99\times$--$1.00\times$ under full GPCPlace. Third, always applying $D$ regardless of $\bar{X} \geq X_{\mathrm{m}}$ overestimates the packing slowdown when L2 contention is low ($\bar{X} < X_{\mathrm{m}}$) and spreads the latency-critical workload across GPCs even when shared-port queueing is severe, leaving it at $0.99\times$--$1.00\times$ of Green Contexts on GPUs with a large $D$ (such as RTX~4090 and RTX~5000~Ada) rather than achieving the $1.04\times$ average speedup of full GPCPlace.

\subsection{Overhead}
\label{sec:eval:overhead}

GPCPlace incurs negligible overhead both offline and at runtime. First, identifying the mapping from TPC mask bits to GPCs (Section~\ref{sec:shape:launch}) is a one-time procedure per target GPU that completes within 3\,s across all 14 GPUs. Second, when profiling workloads before deployment (Section~\ref{sec:shape:decide}), measuring $I$ and $X$ takes about 2--3\,s per background workload, and measuring $L$ and $D$ takes less than 1.5\,s per latency-critical workload. Because these profiled metrics are recorded once and reused across deployments, this one-time cost is negligible for long-running serving systems. Third, at runtime, setting the QMD's TPC mask in the CUDA driver launch callback (Section~\ref{sec:shape:launch}) adds less than $1\,\mu\text{s}$ to the end-to-end kernel launch latency and requires no additional \texttt{ioctl} calls or GPU memory.

\section{Related Work}
\label{sec:related}

\textbf{GPU sharing systems.} Existing systems partition SMs across concurrent workloads to share a GPU, including LLM prefill and decoding~\cite{MW26,SGL25,NX25,CF26,DU25,BU26}, multimodal stages~\cite{HS26,SS25}, multiple models~\cite{MS24,BB25,MO26}, inference and fine-tuning~\cite{HA25}, and real-time tasks~\cite{BV26,ZA21,DY26}. To mitigate the remaining slowdown from shared L2 cache and memory contention~\cite{EL25,SSV25}, existing systems adjust SM counts~\cite{MW26,NX25,BU26,HA25,HS26,BV26,DY26,BB25}, delay or slice co-running kernels~\cite{MO26,GA24,OR24,TA25}, or limit background memory accesses~\cite{BV26}. For example, MuxWise~\cite{MW26} sizes the prefill and decoding partitions with an offline-profiled contention margin, Mosaic~\cite{MO26} admits background kernels based on predicted interference, and Beaver~\cite{BV26} throttles the HBM accesses of co-running ML kernels. GPCPlace instead mitigates contention at shared GPC crossbar ports by deciding whether to spread or pack SMs across GPCs, and keeps each workload's SM count unchanged so that it can be combined with these systems.

\textbf{SM partitioning mechanisms and placement.} Existing SM partitioning mechanisms isolate TPCs without dynamically selecting between spreading and packing across GPCs: CUDA Green Contexts~\cite{GREENCTX} accept only SM counts and spread each workload across GPCs by default, while libsmctrl~\cite{LS23}, nvtaskset~\cite{NT25}, and LithOS~\cite{LO25} expose per-TPC masks, and the first two recommend packing into dedicated GPCs to reduce intra-GPC interference or support thread block clusters. Zhao et al.~\cite{CA24} study assigning SMs across GPCs in a simulator, but always spread concurrent workloads across all GPCs to avoid limiting port bandwidth, overlooking that spreading causes latency-critical requests to queue behind background requests at shared ports. Fan et al.~\cite{DS26} select physical GPCs on large-die GPUs to handle unequal GPC sizes and NUMA locality rather than crossbar port contention. In contrast, GPCPlace profiles both slowdowns on the target GPU and dynamically selects between spreading and packing based on the concurrent workloads and the GPU.

\section{Conclusion}
\label{sec:conclusion}

In this paper, we showed that CUDA Green Contexts always spread workloads across multiple GPCs and cause latency-critical memory requests to queue behind background requests at shared crossbar ports, whereas packing a latency-critical workload into dedicated GPCs avoids this queueing but limits its memory bandwidth through fewer ports, making the faster placement depend on both the concurrent workloads and the GPU hardware. Based on these observations, we presented GPCPlace, a runtime system that profiles the workloads on the target GPU to estimate the shared-port queueing delay and the packing slowdown, dynamically selects between spreading and packing a latency-critical workload, and implements the placement at runtime by hooking kernel launches in the CUDA driver. Across 14 NVIDIA GPUs spanning four architectures on 112 concurrent workload cases, GPCPlace speeds up latency-critical workloads by up to $1.15\times$ on average and up to $2.38\times$ on individual cases over Green Contexts without degrading total system throughput.

\bibliographystyle{IEEEtranS}
\bibliography{refs}

\end{document}